%% file: main.tex
\documentclass{article}
\usepackage{geometry}
\usepackage{graphicx} 
\usepackage[dvipsnames]{xcolor}
\usepackage{xurl}
\usepackage{hyperref}
\usepackage{amsmath}
\usepackage{amssymb} 
\usepackage[table]{xcolor}
\usepackage{booktabs,tabularx,array}
\usepackage[font=footnotesize]{caption}
\usepackage{bm}
\usepackage{float}
\usepackage{caption}
\usepackage{subcaption}

\newcolumntype{L}[1]{>{\raggedright\arraybackslash}p{#1}}
\newcolumntype{Y}{>{\raggedright\arraybackslash}X}

\usepackage[normalem]{ulem}

\title{Tensor-network simulation of quantum transport in many-quantum-dot systems}
\author{Maximilian Streitberger$^{*}$, Marko J. Ran\v{c}i\'{c}$^\dagger$\\
University of Luxembourg, FSTM, Department of Computer Science\\
Campus Belval, 2 avenue de l'Université L-4365 Esch-sur-Alzette\\
\texttt{$^*$maximilian.streitberger@uni.lu}\\ \texttt{$^\dagger$marko.rancic@uni.lu}}
\date{\today}

\begin{document}

\maketitle

\begin{abstract}
Transport through correlated nanoscale systems underpins the operation of quantum-dot and molecular-scale devices, yet accurate simulations of large open quantum systems remain computationally challenging as system size increases. Tensor-network methods offer a promising route past this scaling barrier by efficiently compressing quantum states. Here we extend a tensor-based solver with a jump-counting estimator that enables direct computation of steady-state electron currents from lead-induced tunneling events. We benchmark the resulting currents against the state-of-the-art master-equation solver QmeQ across a range of lead–dot and inter-dot coupling parameters and find quantitative agreement in the tractable regime. Compared with classical approaches, TJM reduces memory requirements and wall-clock time by orders of magnitude, enabling simulations of interacting quantum-dot arrays far beyond the range accessible to density-matrix-based transport solvers and systematic studies of size-dependent nonequilibrium transport in larger arrays. Our approach allow us to model quantum transport in an array of up to fifty (50) quantum dots.
\end{abstract}

\input{introduction}

\input{results}

\input{discussion}
\input{methods}


\input{supplementary_information}

\end{document}

%% file: introduction.tex
\begin{figure}[b]
    \centering
    \includegraphics[width=0.8\linewidth]{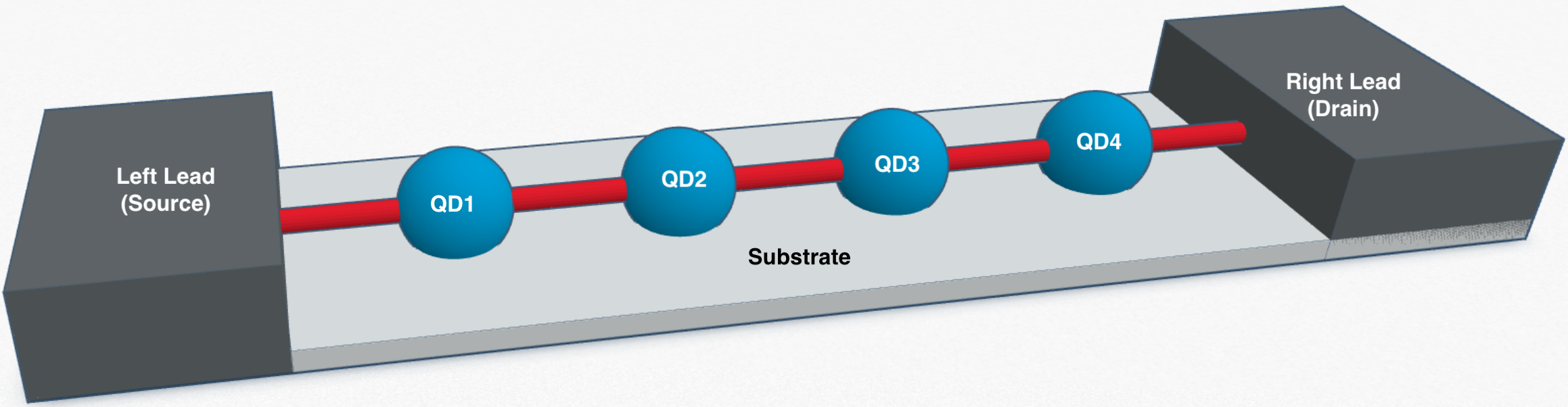}
    \caption{\textbf{Four-quantum-dot device schematic.} Schematic of a linear array of four quantum dots tunnel-coupled in series and connected to source and drain reservoirs.}
    \label{fig:placeholder}
\end{figure}
Charge transport in correlated quantum systems is a central problem in condensed-matter physics and nanoscale device modeling. At the nanoscale, transport is governed by tunneling, phase coherence, level quantization and Coulomb interactions, leading to nontrivial current–voltage characteristics~\cite{hanson_spins_2007,zwanenburg_silicon_2013,Kouwenhoven1997,kastner_single-electron_1992}. Predicting these observables therefore requires computational methods that can treat nonequilibrium quantum dynamics while remaining efficient enough to access experimentally relevant parameter regimes.

Quantum dots provide a natural setting for these challenges because confinement and electron–electron interactions produce a minimal but strongly correlated open-system transport problem~\cite{kastner_artificial_1993}. When coupled to source and drain reservoirs, they realize interacting subsystems exchanging particles and energy with macroscopic environments~\cite{opensystem_2007}. Simulating this dynamics is computationally demanding because it requires resolving the interplay of local interactions, system–lead coupling and nonequilibrium driving as system size increases.

A range of established approaches has been developed for this purpose, but each comes with characteristic trade-offs. Nonequilibrium Green’s function methods provide a powerful framework for steady-state and time-dependent transport and are highly effective in weakly interacting or noninteracting regimes~\cite{nazarov_quantum_2009,jacoboni_theory_2010,Datta_1995,manzano_short_2020}. However, in strongly correlated systems they generally require additional approximations, and their numerical cost increases rapidly when interactions and time dependence are included~\cite{diniz_transport_2026,dorligjav_theoretical_2026}. Quantum master-equation approaches, by contrast, offer an efficient reduced description of open-system dynamics, yet evolving the reduced density matrix still becomes increasingly expensive as the Hilbert-space dimension grows, and the underlying approximations can become restrictive when correlations, coherent dynamics, or more complex system structures are important~\cite{bonnes2014superoperatorsvstrajectoriesmatrix,weimer_simulation_2021}. The main characteristics and limitations of these approaches are summarized in Table~\ref{tab:method_comparison}.

These limitations have motivated stochastic formulations of open quantum dynamics, in which the evolution of a mixed state is unraveled into an ensemble of pure-state realizations, replacing density-matrix propagation with stochastic wavefunction evolution ~\cite{molmer_monte_1993,daley_quantum_2014}. Such methods are commonly known as quantum trajectory methods or Monte Carlo wavefunction (MCWF) approaches. This can substantially reduce memory requirements, especially when the effective state space is large. Combined with tensor-network representations, which compress physically relevant many-body states according to their entanglement structure, this strategy provides a route to simulations that would otherwise be inaccessible~\cite{schollwock_density-matrix_2011,orus_tensor_2019,paeckel_time-evolution_2019}. This combination has recently been realized in the Tensor Jump Method (TJM)~\cite{sander2025}.

TJM provides a computationally promising alternative to conventional density-matrix-based schemes for open quantum dynamics, but it is not yet fully established as a practical tool for quantum transport. Transport studies require not only stable time evolution but also accurate evaluation of observables tied to particle exchange with the reservoirs. Among these, the current is the most fundamental quantity, and a robust strategy for computing it is essential if TJM is to be used beyond proof-of-principle open-system simulations.

Here, we address this gap by extending TJM to enable direct evaluation of particle currents within the tensor-network trajectory formalism. This turns TJM from a general method for dissipative quantum dynamics into a transport-capable framework for interacting nonequilibrium systems. While TJM has previously been used to simulate dissipative quantum dynamics, its application to nonequilibrium quantum transport has remained unexplored. We formulate current extraction in a way that is naturally compatible with stochastic tensor-network evolution, preserving the central computational advantages of the method.

To assess the validity and practical value of the approach, we benchmark it against QmeQ, a widely used reference implementation for quantum-dot transport based on quantum master equations~\cite{qmeq}. By comparing current–voltage characteristics across representative regimes, we test whether the extended TJM reproduces established transport behavior while retaining the flexibility of a trajectory-based tensor-network description. We further analyze runtime and memory consumption to evaluate the computational scaling of the method in practice.

Prior works have addressed transport in interacting open many-body settings and nonequilibrium correlated systems~\cite{prosen_2012, rams_2020}, but these studies considered different transport geometries, observables, or methodological frameworks than the tensor-network trajectory approach developed here, and especially outside the many-quantum dot problem studied here.

Other relevant studies have treated transport in more limited or adjacent settings, such as quantum-impurity models, discretized leads, or more general open-system tensor-network formulations~\cite{schwarz_2016, lotem_2020, chen_2024, thoenniss_2023}, without addressing steady-state charge transport through extended interacting many-body quantum-dot arrays in the form considered in this work.

We demonstrate that TJM can be equipped with a reliable current estimator and thereby turned into a practical computational workflow for nonequilibrium transport. More broadly, they indicate that tensor-network quantum trajectories provide a viable and extensible route to transport simulations in correlated open quantum systems, particularly in regimes where direct density-matrix propagation becomes impractical.

\begin{table}[t]
\centering
\scriptsize
\setlength{\tabcolsep}{2pt}
\renewcommand{\arraystretch}{1.2}
\begin{tabularx}{\linewidth}{L{2.8cm} Y Y Y}
\toprule
& \textbf{NEGF} & \textbf{Master equation} & \textbf{TJM (this work)} \\
\midrule
\textbf{Type}
& Deterministic
& Deterministic
& Stochastic \\
\arrayrulecolor{gray!25}
\midrule
\textbf{State}
& Green's functions
& Reduced density matrix
& Pure-state trajectories \\
\midrule
\textbf{Time dependence}
& Yes
& Limited
& Yes \\
\midrule
\textbf{Interactions}
& Exact for noninteracting systems; approximate otherwise
& Effective in reduced open-system settings
& Interacting many-body dynamics via trajectories \\
\midrule
\textbf{Current access}
& Direct
& Direct
& Jump-counting estimator \\
\midrule
\textbf{Scaling}
& Discretization, self-consistency, correlated self-energies
& Density-matrix dimension
& Bond dimension and trajectory count \\
\midrule
\textbf{Best for}
& Noninteracting or weakly correlated transport
& Reduced open-system descriptions
& Strongly correlated open systems \\
\midrule
\textbf{Limitation}
& Strong correlations depend on interaction approximation
& Often relies on weak-coupling / Markov assumptions
& Depends on tensor-network accuracy and TDVP evolution \\
\arrayrulecolor{black}
\bottomrule
\end{tabularx}
\caption{Qualitative comparison of common approaches for nonequilibrium transport in open quantum systems. Representative references for the methods shown here include NEGF \cite{ryndyk_2009}, master-equation approaches \cite{manzano_short_2020}, and TJM \cite{sander2025}.}
\label{tab:method_comparison}
\end{table}

%% file: results.tex
\section*{Results}\label{sec:results}

\subsection*{Extending TJM to compute transport currents}
We extend the tensor jump method (TJM) by introducing jump counters that record the individual applications of lead-induced jump operators during the stochastic trajectory evolution. These counters allow us to determine the net number of charges transferred through each lead–dot channel over time, from which the particle current can be estimated. The jump counts are ensemble-averaged over trajectories to obtain the steady-state current. Implementation details and the precise current definition are provided in the Methods section.

\subsection*{Validation against master-equation benchmarks}
We benchmark TJM against the Lindblad master-equation implementation in QmeQ in the parameter regime where direct density-matrix simulations remain tractable. Because the Liouville-space dimension grows rapidly with system size, these reference calculations were limited to arrays of up to four quantum dots. For each benchmark point, we compared the steady-state current obtained from TJM and QmeQ and quantified their pointwise deviation as
\begin{equation}
    \Delta I = \left| I_{\mathrm{TJM}} - I_{\mathrm{QmeQ}} \right|.
\end{equation}
To keep the main text focused, we highlight the two most informative sweeps: the lead--dot coupling $\Gamma$ and the inter-dot hybridization $\Omega$. The remaining parameter sweeps, together with the full error analysis, are provided in the Supplementary Information.

\begin{figure}
    \centering
    \includegraphics[width=1\linewidth]{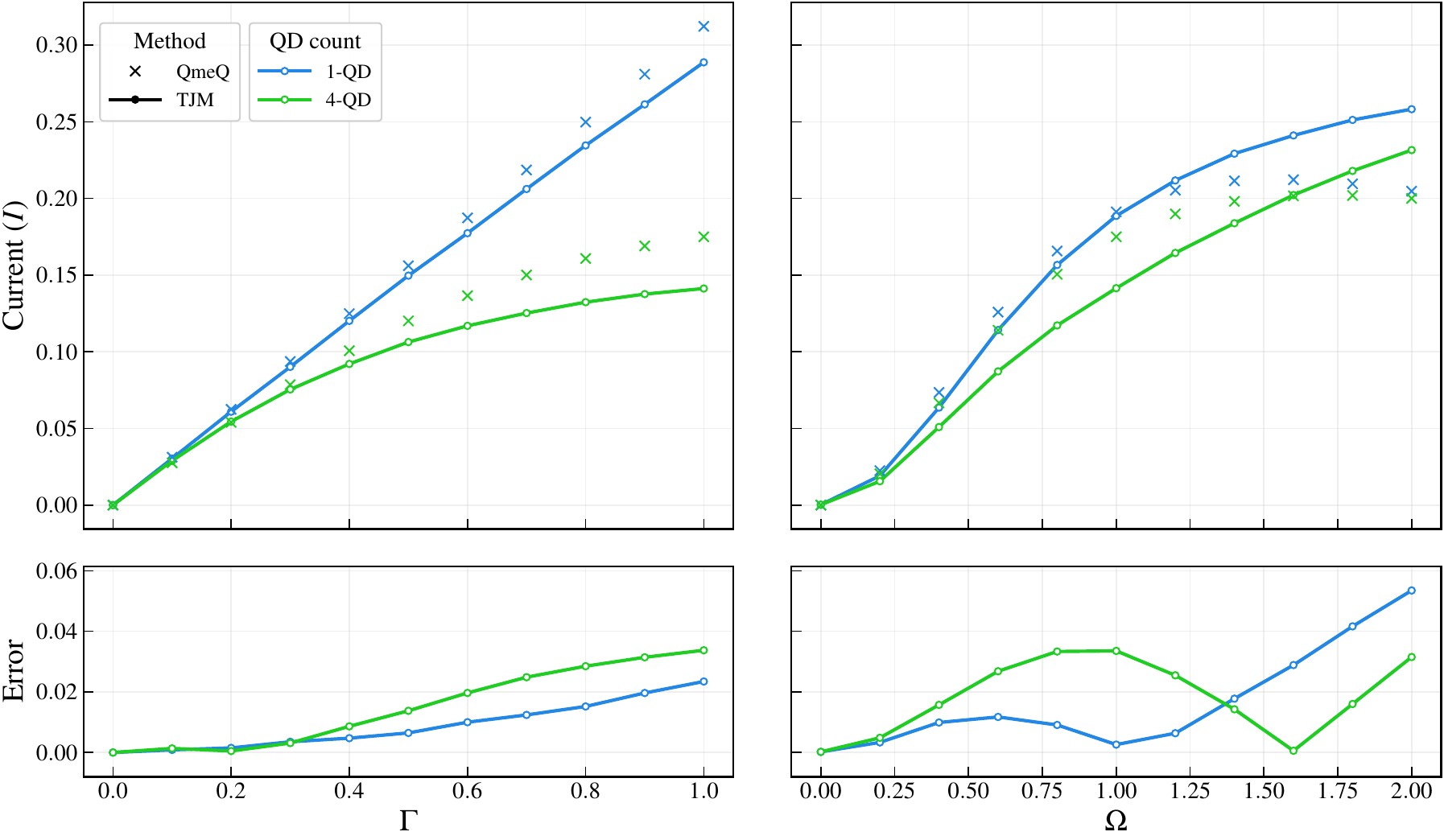}
    \caption{\textbf{Current and method discrepancy as functions of lead--dot and inter-dot coupling.} Top panels show the current as a function of the lead--dot coupling, $\Gamma$ (left), and the inter-dot coupling, $\Omega$ (right), for single-quantum-dot (1-QD, blue) and four-quantum-dot (4-QD, green) systems. QmeQ results are shown as crosses, and TJM results as circles connected by solid lines. Bottom panels show the corresponding absolute error between the two methods for the same parameter sweeps. Agreement is strong across the full parameter range, with discrepancies remaining small and becoming more pronounced only at larger coupling strengths.}
    \label{fig:qmeq_tjm_comparison}
\end{figure}
\vspace{1em}
\noindent\textbf{Lead--dot coupling.} Increasing the lead--dot coupling, $\Gamma$, produces a monotonic increase in the stationary current in both the smallest and largest benchmark systems (Fig.~\ref{fig:qmeq_tjm_comparison}). Across the full sweep, TJM closely follows the QmeQ reference, reproducing both the overall growth in current and the relative separation between the two device sizes.

For the smaller system, the current rises from zero to $0.312$ in QmeQ, while TJM reaches $0.289$ at the largest $\Gamma$. The corresponding absolute deviation remains small throughout the sweep and increases only gradually, reaching $2.34\times10^{-2}$ at the largest coupling. The same qualitative behavior is observed in the larger system, for which the current increases less strongly, from zero to $0.175$ in QmeQ and $0.141$ in TJM, with a maximum deviation of $3.38\times10^{-2}$. In both cases, the TJM and QmeQ curves remain close over the entire parameter range, with no indication of a qualitative breakdown as the coupling to the leads is increased.

These results show that stronger lead coupling primarily amplifies the transport response while leaving the agreement between the two approaches largely intact. Within the tested range, TJM therefore captures the lead-coupling dependence of the stationary current, from the smallest system considered to the largest system accessible to direct master-equation benchmarking.

\vspace{1em}
\noindent\textbf{Inter-dot hybridization.} Varying the inter-dot coupling, $\Omega$, provides a stronger test of the model because it directly enhances coherent hybridization within the array (Fig.~\ref{fig:qmeq_tjm_comparison}). In contrast to the lead--dot sweep, the agreement between TJM and QmeQ remains good at weak-to-intermediate $\Omega$, with systematic deviations emerging as hybridization increases.

In the smaller multi-dot system, both methods initially predict a strong rise in current with increasing $\Omega$. However, the QmeQ current reaches a maximum at intermediate coupling ($\sim 0.21$) and then saturates slightly, whereas the TJM current continues to increase up to the largest $\Omega$ considered ($\sim 0.26$). As a result, the absolute deviation grows steadily across the sweep and is largest at strong hybridization ($\sim 5.3\times10^{-2}$).

A related but more structured pattern is seen in the larger system. Here too, TJM reproduces the initial increase in current, but the discrepancy is non-monotonic: it becomes most pronounced at intermediate coupling, decreases again when the two curves nearly coincide, and then rises once more at the strongest hybridization. This indicates that the breakdown is not simply set by the magnitude of $\Omega$, but by how strongly coherent inter-dot mixing reshapes the transport pathway in different parts of parameter space.

Together, these results identify inter-dot hybridization as the main regime in which TJM deviates from the QmeQ reference. The method remains accurate at weak coupling, but strong coherent mixing introduces the largest and most systematic discrepancies observed in the benchmark set.

The growing discrepancy between TJM and QmeQ at large inter-dot hybridization is consistent with the known breakdown of local Lindblad master equations outside the weak inter-system-coupling regime. In this regime, the benchmark itself is expected to yield incorrect stationary currents, so the observed divergence should not be interpreted solely as a limitation of TJM. Instead, it delineates the parameter regime in which agreement with a local master-equation reference can be expected~\cite{cattaneo_local_2019}.

\vspace{1em}
\noindent\textbf{Additional parameters.} Additional parameter sweeps (onsite energy, temperature and Coulomb interaction) show similar agreement and are provided in the Supplementary Information (see S1 - S3).

\subsection*{Wall-clock-time and memory scaling}

\begin{figure*}[t]
    \centering
    \begin{minipage}{0.49\textwidth}
        \centering
        \subfloat[\label{fig:runtime_scaling}]{
            \includegraphics[width=0.97\linewidth]{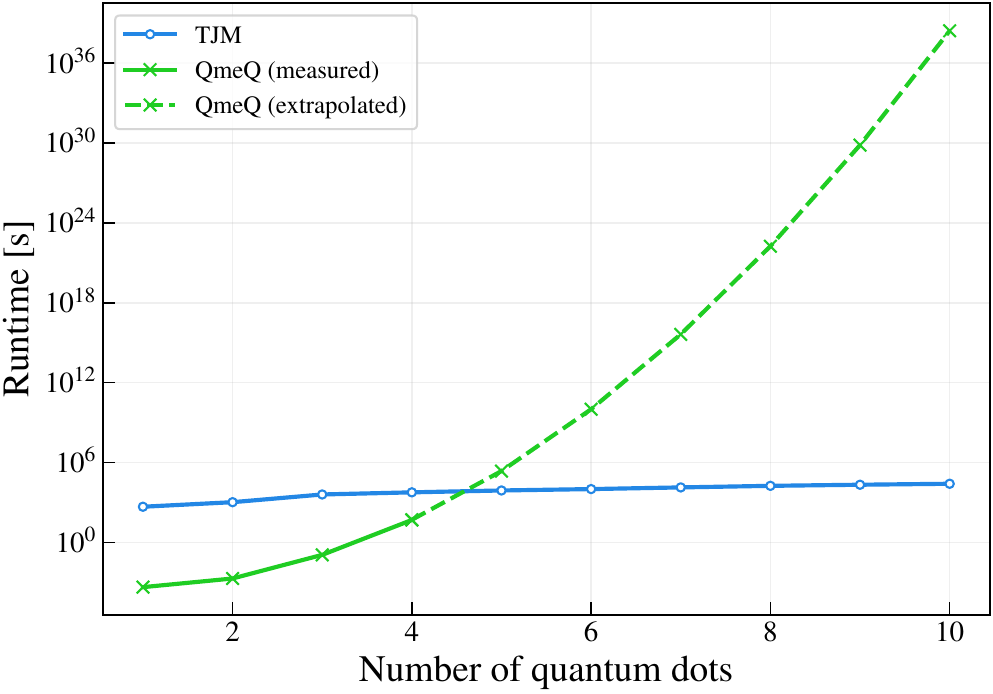}
        }
    \end{minipage}\hfill
    \begin{minipage}{0.49\textwidth}
        \centering
        \subfloat[\label{fig:memory_scaling}]{
            \includegraphics[width=0.97\linewidth]{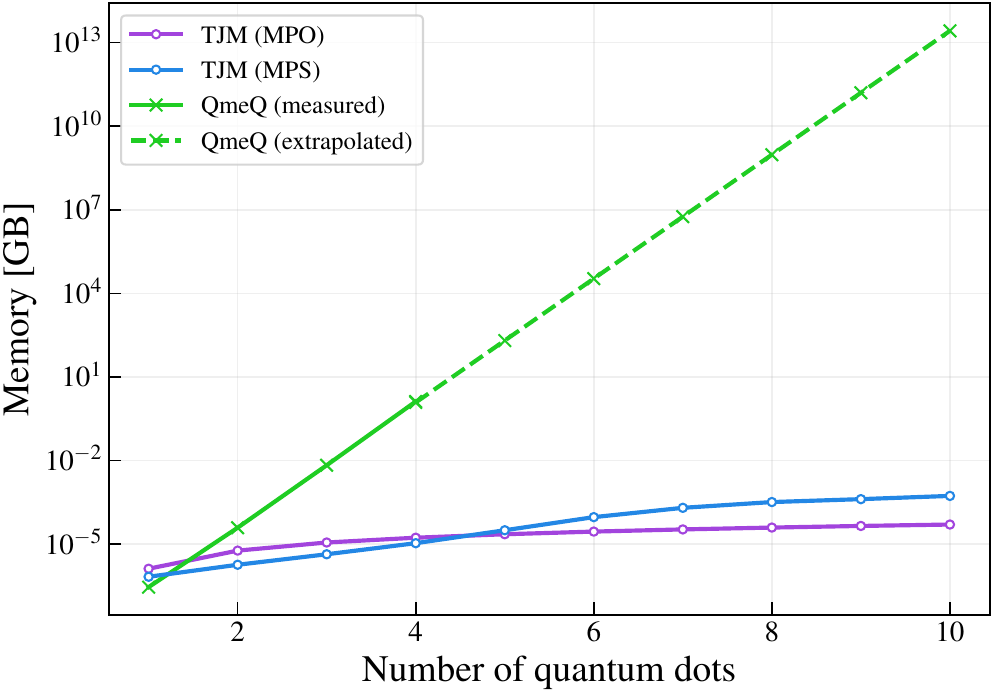}
        }
    \end{minipage}
    \caption{\textbf{Runtime and memory scaling of TJM and QmeQ with system size.} (a): Wall-clock time as a function of the number of quantum dots for TJM and QmeQ. Although QmeQ is faster for small systems, its computational cost increases much more rapidly with system size. TJM becomes advantageous beyond the crossover regime and remains the more efficient approach for larger systems. (b): Memory requirement as a function of the number of quantum dots. For TJM, the memory footprint of the stored MPS (blue) and MPO (purple) remains small because both the state and the Hamiltonian are represented in tensor-network form, whereas the memory required by QmeQ increases steeply with system size. The MPS memory grows faster than the MPO memory because its bond dimension can increase up to $\chi_{\max}$.}
    \label{fig:memory_and_runtime}
\end{figure*}

\noindent\textbf{Wall-clock-time scaling.}
The wall-clock-time comparison shows a different behavior at small system sizes. In the directly tractable regime, QmeQ is faster because TJM incurs the additional cost of explicit time evolution and stochastic averaging. For example, at $N=4$, QmeQ completes in $4.81\times10^{1}\,\mathrm{s}$, compared with $5.73\times10^{3}\,\mathrm{s}$ for TJM. However, the TJM runtime grows only moderately over the full 1--10 dot range, whereas the projected QmeQ runtime increases much more steeply. The crossover occurs by $N=6$, where TJM is already faster, and this advantage then grows rapidly with system size (Fig.~\ref{fig:runtime_scaling}). Together with the memory analysis, these results identify TJM as the computationally favorable approach beyond the small-system regime.

\vspace{1em}
\noindent\textbf{Memory scaling.}
We next compared the memory requirements of TJM and QmeQ as a function of system size. TJM maintains a very small memory footprint across the full 1--10 dot range because both the Hamiltonian and the state are stored in tensor-network form. Its memory requirement scales with the number of trajectories, system size and maximal bond dimension. By contrast, the memory requirement of the Lindblad solver grows combinatorially with the number of single-particle states because the reduced density matrix is represented in the full many-body basis. Even at $N=4$, QmeQ already requires $1.33\,\mathrm{GB}$, whereas TJM requires only $3.37\times10^{-5}\,\mathrm{GB}$, a reduction of roughly five orders of magnitude. This disparity widens further with increasing $N$, and by $N=10$ the projected QmeQ memory requirement is prohibitively large (Fig.~\ref{fig:memory_scaling}). These results show that tensor-network trajectories avoid the unfavorable memory growth of the dense master-equation formulation and therefore remain tractable at substantially larger system sizes. Figure~\ref{fig:memory_scaling} separates the TJM memory cost into the MPS state and MPO Hamiltonian contributions. The MPS contribution grows faster, since the MPO scales approximately linearly with system size, whereas the MPS bond dimension can increase up to the maximum allowed value ($\chi_{\max}$) as inter-dot correlations build up.

\subsection*{Transport calculations beyond the direct benchmark regime}
\begin{figure}
    \centering
    \includegraphics[width=0.65\linewidth]{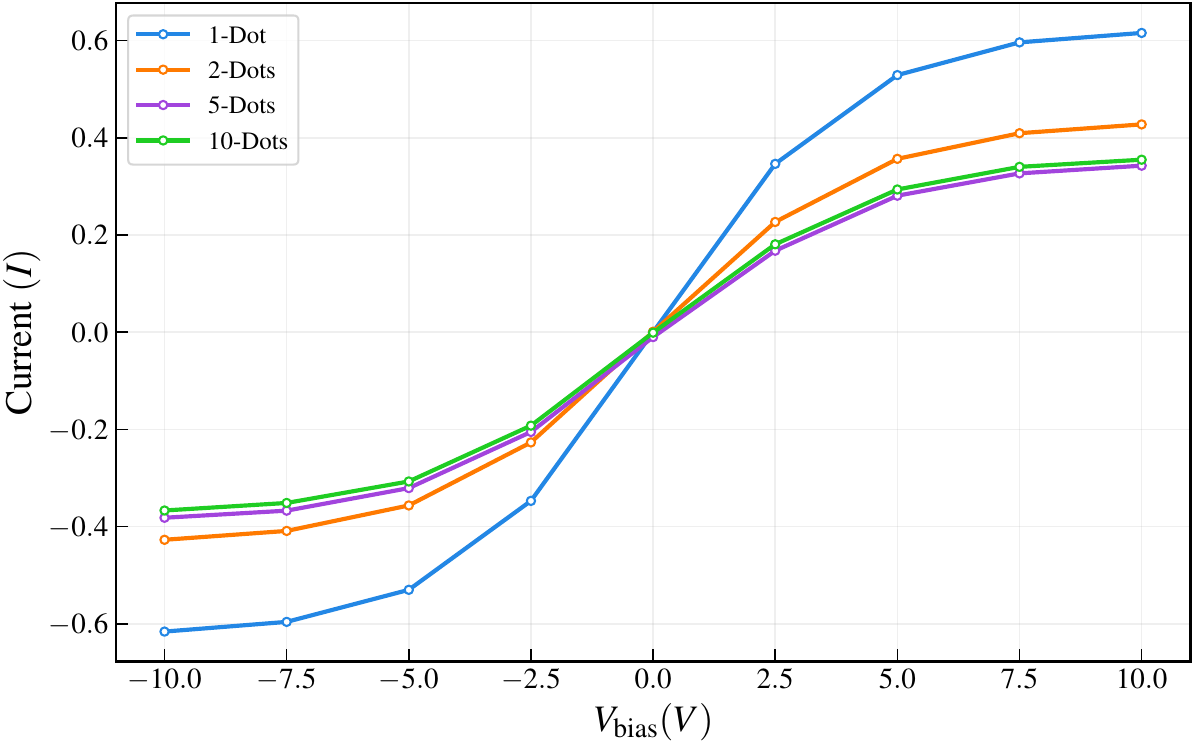}
    \caption{\textbf{Source--Drain bias sweep.} Here the left lead is acting as the source and the right lead as the drain. The potential difference corresponds to the bias $V_\text{bias}=\mu_L - \mu_R$. We can see that the current decreases systematically with increasing array length.}
    \label{fig:v_sd_sweep}
\end{figure}
Having established the validity of the current estimator in the tractable regime, we next used TJM to study transport in larger arrays beyond the range accessible to direct master-equation benchmarking. Figure~\ref{fig:v_sd_sweep} shows the resulting current--voltage characteristics for systems containing up to ten quantum dots. As the array length increases, the current is progressively suppressed across the full bias window, while the overall nonlinear transport structure is retained. TJM therefore not only extends the accessible system size, but also resolves physically meaningful size-dependent transport trends in regimes where dense many-body reference calculations become impractical.

\begin{figure*}[t]
    \centering
    \begin{minipage}{0.49\textwidth}
        \centering
        \subfloat[\label{fig:50_dots_current}]{
            \includegraphics[width=0.97\linewidth]{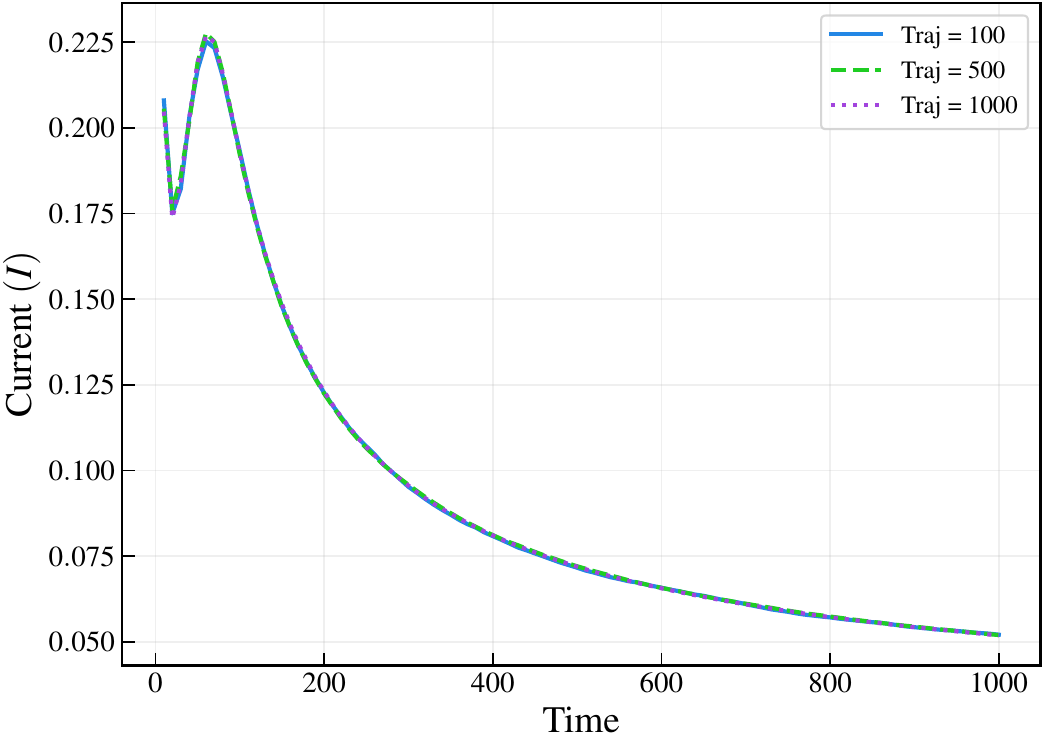}
        }
    \end{minipage}\hfill
    \begin{minipage}{0.49\textwidth}
        \centering
        \subfloat[\label{fig:50_dots_mismatch}]{
            \includegraphics[width=0.97\linewidth]{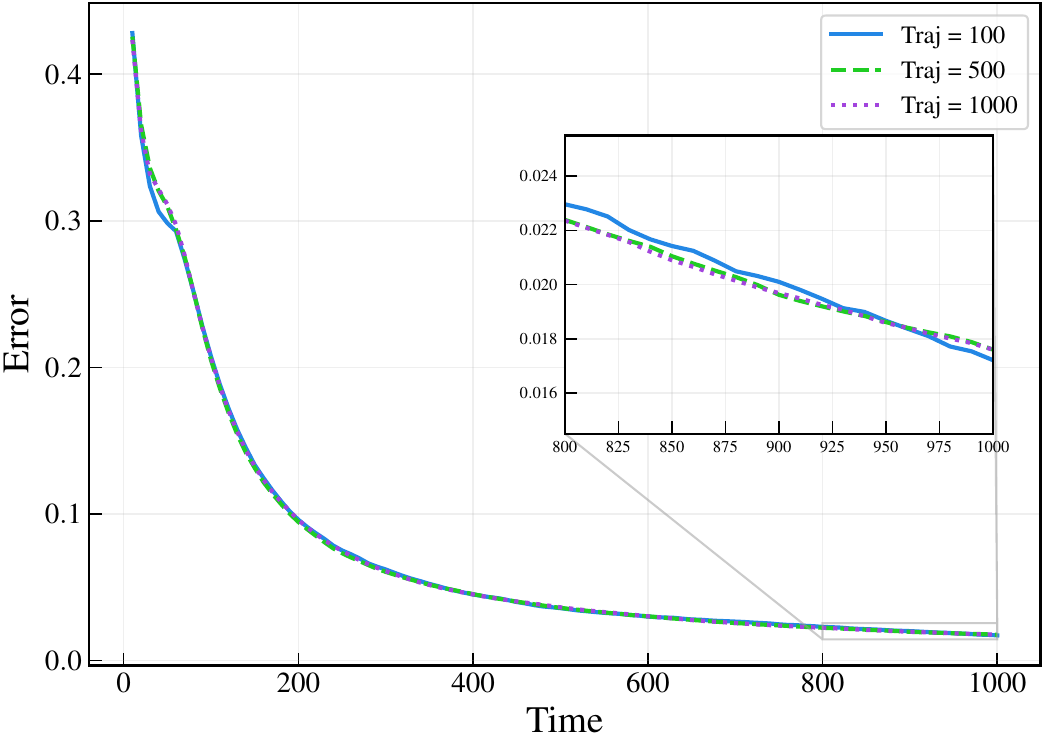}
        }
    \end{minipage}
    \par\vspace{0.5cm}
    \begin{minipage}{0.49\textwidth}
        \centering
        \subfloat[\label{fig:runtime_vs_trajectories}]{
            \includegraphics[width=0.97\linewidth]{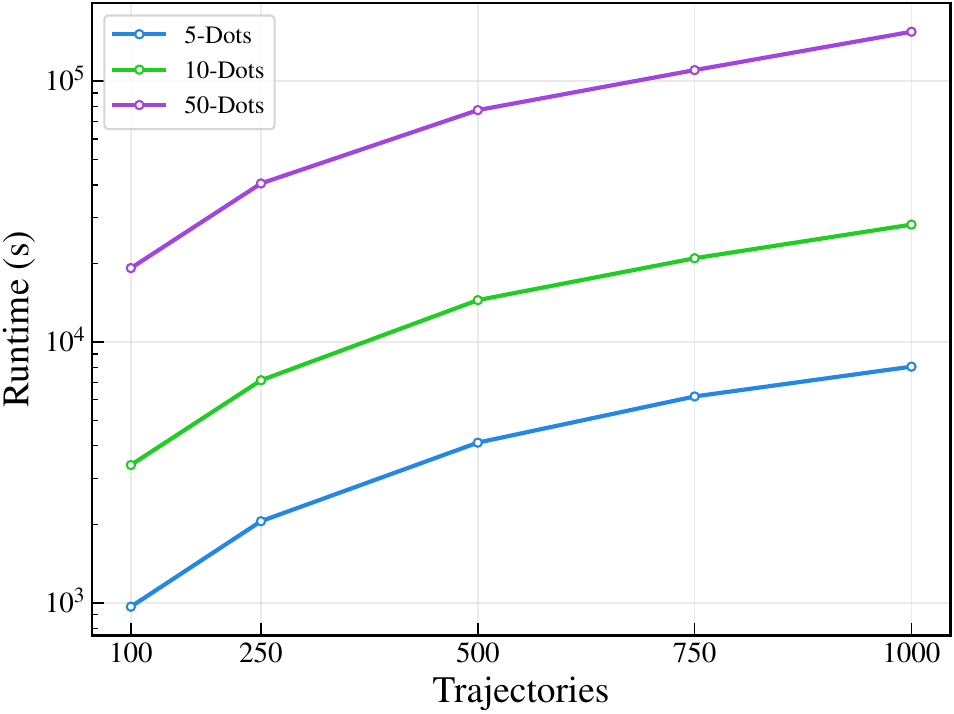}
        }
    \end{minipage}
    \caption{Current dynamics and runtime scaling beyond the direct benchmark regime. (a) Current dynamics in a 50-dot array computed with TJM for different ensemble sizes. Increasing the number of trajectories reduces statistical fluctuations, but the current traces remain closely aligned over the plotted interval and do not approach a clear stationary plateau. This indicates that slow relaxation towards the steady state, rather than insufficient sampling, is the dominant limitation in this regime. (b) Left--right lead-current mismatch for the 50-dot array, which decreases during the time evolution and remains small at late times, providing an internal consistency check. (c) Wall-clock time versus number of trajectories for 5-, 10-, and 50-dot arrays. Runtime increases with both ensemble size and system size, highlighting the growing computational cost of large-array TJM simulations and complementing the memory-scaling analysis.}
\end{figure*}

To probe the behavior at still larger scales, we additionally simulated the time-dependent current in a 50-dot array at a source--drain bias of $V_\text{bias}=10$ (Fig.~\ref{fig:50_dots_current}). Within the simulated time window, the current starts to approach a stationary plateau after $t=1000$, indicating that relaxation towards the steady state becomes markedly slower as system size increases. By contrast, increasing the number of trajectories produces only comparatively modest changes in the current trace over the same interval. This indicates that the dominant bottleneck in this regime is slow relaxation rather than trajectory undersampling. As an internal consistency check, the mismatch between the left and right lead currents decreases during the propagation and remains small at late times, supporting the numerical stability of the calculation even when full steady-state convergence is not reached (see Fig.~\ref{fig:50_dots_mismatch}).

The associated runtime increases sharply with both system size and ensemble size (Fig.~\ref{fig:runtime_vs_trajectories}). Taken together, these results show that TJM substantially enlarges the accessible system-size range for nonequilibrium transport calculations while identifying long-time relaxation as the main remaining bottleneck in large arrays.


%% file: discussion.tex
\section*{Discussion}
We have shown that the Tensor Jump Method (TJM) can be extended to compute steady-state electron currents in open quantum-dot systems by extracting the current directly from lead-resolved stochastic jump events. Benchmarking against the Lindblad master-equation solver in QmeQ shows that the resulting jump-count estimator is quantitatively accurate across a broad transport regime. This turns the quantum-jump dynamics itself into a practical measurement protocol for nonequilibrium transport while avoiding explicit density-matrix evolution.

TJM reproduces the reference currents well over sweeps in lead--dot coupling, temperature, onsite energy and Coulomb interaction, indicating that the estimator captures both the correct trends and current magnitudes in the regime where direct comparison is possible. The largest deviations arise at strong inter-dot hybridization, where the discrepancy increases with coupling strength and system size. This behavior is consistent with the known limitations of local master-equation descriptions outside the weak inter-system-coupling regime and therefore helps define the parameter range in which the present approach is expected to be most reliable.

The computational significance of this result lies in the scaling benefits of TJM. Although the present implementation is slower than QmeQ in the small-system regime where both methods remain tractable, its memory requirements scale far more favorably. TJM therefore becomes advantageous precisely in the regime where density-matrix-based approaches become impractical. In this sense, the method extends transport calculations into interacting many-body regimes that are difficult to access with conventional dense-kernel solvers.

Beyond the directly benchmarkable regime, TJM also resolves physically meaningful transport trends in larger quantum-dot arrays. The current--voltage characteristics for systems containing up to ten dots show a systematic suppression of current with increasing array length while preserving the overall nonlinear transport structure, demonstrating that the method is not only computationally scalable but also capable of accessing size-dependent nonequilibrium transport behavior in interacting open systems. At still larger scales, the 50-dot simulations show that the dominant limitation shifts from memory consumption and trajectory sampling to slow relaxation towards the steady state. This identifies long-time convergence, rather than state representation, as the main remaining bottleneck for transport calculations in large arrays.

The most immediate opportunities for improvement lie in reducing the cost of long-time evolution, for example through GPU-accelerated tensor operations, lower-level implementations of the main computational bottlenecks and improved parallelization across trajectories. More broadly, our results suggest that tensor-network trajectory methods can provide a scalable route to transport simulations in larger and more strongly interacting open quantum systems than are readily accessible with conventional master-equation solvers.

%% file: methods.tex
\section*{Methods}
\label{sec:methods}

\subsection*{Tensor jump method}
In TJM, the stochastic time evolution of a single trajectory consists of three parts: (i) a dynamic TDVP step $(\mathcal{U})$, (ii) a dissipative contraction $(\mathcal{D})$, and (iii) a stochastic jump process $(\mathcal{J})$. We start from an initial state $|\Psi(0)\rangle$ at $t=0$, represented as an MPS, and evolve it under a system Hamiltonian encoded as an MPO up to a final time $t_{\max} = n\delta t$. 

The coherent (Hamiltonian) evolution of the MPS is performed using the time-dependent variational principle (TDVP). TJM applies the evolution dynamically: it starts with two-site TDVP (2TDVP), which permits adaptive growth of the MPS bond dimension (and thus increasing entanglement) up to a predefined maximum. Once this maximum is reached, the algorithm switches to one-site TDVP (1TDVP) to continue the time evolution within a fixed bond dimension, preventing further growth of the variational manifold. 

The dissipative part is described by a set of jump operators, analogous to the MCWF approach $\{L_m\}_{m=1}^k$. The time evolution of the state can be expressed as
\begin{equation}
    U(t_{\max}) = \prod_{i=0}^{n} \mathcal{F}_{n-i}[\delta t],
\end{equation}
\noindent where the propagator $U$ consists of $n$ subfunctions corresponding to each time step:
\begin{equation}
    \mathcal{F}_j[\delta t] = 
    \begin{cases}
        \mathcal{J}_\epsilon[\delta t] \  \mathcal{D}[\frac{\delta t}{2}] \ \mathcal{U}[\delta t]   & \quad j=n,\\
        \mathcal{J}_\epsilon[\delta t] \ \mathcal{D}[\delta t] \ \mathcal{U}[\delta t]              & \quad 0 < j < n,\\
        \mathcal{J}_\epsilon[\delta t] \ \mathcal{D}[\frac{\delta t}{2}]                            & \quad  j=0.\\
    \end{cases}
\end{equation}

Internally, TJM propagates an auxiliary “sampling MPS” $|\Phi\rangle$ with the maps $\mathcal{F}_j[\delta t]$, and the physical state $|\Psi\rangle$ can be reconstructed from $|\Phi\rangle$ at any time step by applying the final operator $\mathcal{F}_n[\delta t]$.

The sequence of maps $\{\mathcal{F}_j[\delta t]\}_{j=0}^{n}$ together with a specific realization of the stochastic jumps defines a single stochastic trajectory. Physical observables are obtained by averaging over $N$ such independent trajectories, in full analogy with the MCWF approach.

\subsection*{Open-system model}

We consider a device consisting of several quantum dots coupled to two metallic leads. The system is described by
\begin{equation}
    H = H_{\text{leads}} + H_{\text{tunneling}} + H_{\text{dot}},
    \label{eq:model}
\end{equation}
where $H_{\text{leads}}$ describes the leads, $H_{\text{dot}}$ the system of tunnel coupled quantum dots, isolated from the leads, and $H_{\text{tunneling}}$ the lead-dot tunneling. We can write the system in second-quantized form as
\begin{equation}
    \begin{split}
        H =\;& \sum_{\alpha k} \epsilon_{\alpha k}\, c^\dagger_{\alpha k} c_{\alpha k}
        + \sum_{\alpha k,i}\left(t_{\alpha k,i}\, d_i^\dagger c_{\alpha k} + \text{H.c.}\right) \\
        &+ \sum_i \epsilon_i\, d^\dagger_i d_i
        + \sum_{i \ne j} \Omega_{ij}\, d_i^\dagger d_j\\
        &+ \sum_{ij} U_{ijji}\, d_i^\dagger d_j^\dagger d_j d_i,
        \quad \text{with } i<j.
    \end{split}
    \label{eq:model_full}
\end{equation}
We assume that the leads are thermal reservoirs characterized by Fermi-Dirac distributions $f_\alpha(E)$. Moreover, we work in the wide-band limit, i.e., a constant density of states and energy-independent tunneling amplitudes.

Since we evolve the reduced dot state using the Lindblad master equation, only $H_{\text{dot}}$ enters the coherent part of the dynamics. This Hamiltonian is then converted into an MPO.

The effect of the leads enters the Lindblad dynamics through the jump operators (local Lindblad operators), whose factors are determined by the tunneling rates and the lead occupations. We use injection and extraction operators
\begin{equation}
    \begin{split}
        L_{\alpha,i}^{\text{in}}  &= \sqrt{\Gamma_{\alpha i}\, f_\alpha(E_i)}\; d_i^\dagger,\\
        L_{\alpha,i}^{\text{out}} &= \sqrt{\Gamma_{\alpha i}\, \bigl[1-f_\alpha(E_i)\bigr]}\; d_i,
    \end{split}
\end{equation}
where $\Gamma_{\alpha i}$ is the tunneling rate and $E_i$ denotes the relevant transition energy. In particular, $E_i$ may include interaction-induced shifts (addition energies) in the Coulomb blockade regime. Here the transition energy $E_i$ includes the onsite interaction shift: if the opposite-spin level on dot $i$ is occupied, we use the addition energy $E_i = \epsilon_i + U_i$ and otherwise $E_i = \epsilon_i$. Since we retain onsite interactions only, this shift depends solely on the local opposite-spin occupancy.

The tunneling rates are related to the tunneling amplitudes by $\Gamma_{\alpha i} = 2\pi \nu_F |t_{\alpha i}|^2$~\cite{qmeq}.

Under the Jordan–Wigner mapping, a local fermionic annihilation/creation operator in Eq. \ref{eq:model_full} on site $l$ is represented as 
\begin{equation}
    \begin{split}
        d_l &\Leftrightarrow Z^{\otimes (l-1)}\otimes\sigma^+\otimes I^{\otimes (L-l)},\\
        d^\dagger_l &\Leftrightarrow Z^{\otimes (l-1)}\otimes\sigma^-\otimes I^{\otimes (L-l)},
    \end{split}
\end{equation}

\noindent where the product of $Z$ operators encodes the fermionic parity of all sites to the left and ensures the correct anticommutation relations. For a chain of length $L$ and a local operator acting on site $l$ (with $1\le l\le L$), we represent the embedded jump operator as
\begin{equation}
    \begin{split}
        L_m &= Z^{\otimes(l-1)}\otimes L_m^{[l]} \otimes I^{\otimes(L-l)},\\
        L_m^\dagger &= Z^{\otimes(l-1)}\otimes (L_m^\dagger)^{[l]} \otimes I^{\otimes(L-l)}.
    \end{split}
\end{equation}
\noindent Using $Z^\dagger = Z$, $ZZ=I$, and $I^\dagger=I$, the Jordan-Wigner strings cancel in $L_m^\dagger L_m$,
\begin{equation}
    \begin{split}
        L_m^\dagger L_m
        &= (Z^\dagger Z)^{\otimes(l-1)}\otimes (L_m^\dagger L_m)^{[l]} \otimes (I^\dagger I)^{\otimes(L-l)} \\
        &= I^{\otimes(l-1)}\otimes (L_m^\dagger L_m)^{[l]} \otimes I^{\otimes(L-l)}.
    \end{split}
\end{equation}
Thus, the non-Hermitian contribution involving $L_m^\dagger L_m$ is unaffected by the Jordan-Wigner strings, while applying the jump operator $L_m$ in a trajectory still requires the full $Z$ string.

\subsection*{Particle current definition}
We define the (lead-resolved) particle current as the net number of electrons transferred per unit time, which corresponds to the application of a jump operation. Let $\alpha\in\{L,R\}$ label the lead and $\sigma\in\{\uparrow,\downarrow\}$ label the spin channel. During a total simulation time $t_{\max}$, we record the number of quantum-jump events corresponding to electron injection into the system, $n_{\alpha,\sigma}^{\mathrm{in}}$, and extraction from the system, $n_{\alpha,\sigma}^{\mathrm{out}}$, through lead channel $(\alpha,\sigma)$. This recording is done after the full time step dissipation evolution. Within the quantum trajectories approach, all jump counts are ensemble-averaged over trajectories.

The net transferred particle number through channel $(\alpha,\sigma)$ is
\begin{equation}
    N_{\alpha,\sigma} = n_{\alpha,\sigma}^{\mathrm{in}} - n_{\alpha,\sigma}^{\mathrm{out}} \, ,
    \label{eq:net_counts}
\end{equation}
and the corresponding channel current is
\begin{equation}
    I_{\alpha,\sigma} = \frac{N_{\alpha,\sigma}}{t_{\max}} \, .
    \label{eq:channel_current}
\end{equation}
The spin-summed current for lead $\alpha$ is then
\begin{equation}
    I_\alpha = \sum_{\sigma} I_{\alpha,\sigma}
            = \frac{1}{t_{\max}}\sum_{\sigma} N_{\alpha,\sigma} \, .
    \label{eq:lead_current}
\end{equation}

In a stationary regime, particle conservation implies $I_L = -I_R$ (for the sign convention where $I_\alpha>0$ denotes net flow from lead $\alpha$ into the system). For finite simulation times $t_{\max}$, deviations from $I_L=-I_R$ can occur due to statistical fluctuations. We therefore report the symmetrized (transport) current,
\begin{equation}
    I_{\mathrm{through}} = \frac{I_L - I_R}{2} \, ,
    \label{eq:through_current}
\end{equation}
which reduces variance and approaches the ideal steady-state value as the total simulation time ($t_{\max}$) increases.

\subsection*{Numerical setup and benchmarking protocol}

For the benchmark calculations, all simulations were run up to a total time $t_{\max}=1000$ with a time step $\delta t=0.1$. Unless stated otherwise, the lead chemical potentials were fixed at $\mu_L=1$ and $\mu_R=-1$. TJM results were obtained from $N_{\mathrm{traj}}=1000$ stochastic trajectories with a maximum bond dimension $\chi_{\max}=20$. The system was placed symmetrically between the two leads, with no magnetic field (no Zeeman splitting) and spin orbital energies $\epsilon_i=0$. 

Reference calculations were performed with the QmeQ package using the Lindblad kernel and a bandwidth of $d_{\mathrm{band}}=60$. Because the Liouville-space dimension grows rapidly with system size, QmeQ benchmarks were restricted to systems of up to four quantum dots. We compared TJM and QmeQ across parameter sweeps in lead--dot coupling, onsite energy, Coulomb interaction, temperature and inter-dot coupling. For each sweep we evaluated both the current and the pointwise absolute deviation,
\begin{equation}
\Delta I = \left| I_{\mathrm{TJM}} - I_{\mathrm{QmeQ}} \right|.
\end{equation}

\subsection*{Trajectory convergence}
Convergence with respect to trajectory number was assessed by comparing current traces for $N_{\mathrm{traj}} = 100$, $500$ and $1000$. Increasing the number of trajectories reduces statistical fluctuations, while the $500$- and $1000$-trajectory results remain very close for representative $5$-dot and $10$-dot systems, indicating that $N_{\mathrm{traj}} = 1000$ is sufficient for the reported calculations.

To quantify this, we evaluated the difference between the current traces obtained with $500$ and $1000$ trajectories. For representative $5$-dot and $10$-dot systems, the late-time difference between the two estimates remains below $10^{-3}$, as shown in Supplementary Fig. S4. Because the steady-state current is extracted from the long-time behavior of the trajectories, agreement at late times provides the most relevant convergence criterion.

\subsection*{Bond-dimension convergence}
We chose the maximum bond dimension, $\chi_\mathrm{max}$, in the TJM based on convergence tests. Specifically, we compared the current obtained from a lead–dot parameter sweep for systems with one and four quantum dots using $\chi_\mathrm{max}\in{10,20,40,60}$. The resulting currents were effectively unchanged across this range, indicating convergence with respect to $\chi_\mathrm{max}$ for systems up to four quantum dots. Thus, the bond dimension was not a limiting factor in our simulations, unlike the timestep size (see next section). The results are shown in Supplementary Fig. S5.

This observation should not be interpreted as evidence that the smallest bond dimension is sufficient for all system sizes. Because the bond dimension controls the amount of entanglement retained in the tensor-network representation, larger systems may require larger $\chi_\mathrm{max}$ values. Establishing convergence more generally would therefore require analogous tests for larger systems, and, where possible, validation against experiment.

\subsection*{Timestep choice}
The timestep $\delta t$ used in the TDVP evolution was chosen based on a convergence analysis. We compared the current obtained for a single quantum dot and a four-dot array over a range of timestep sizes $\delta t \in \{0.05, 0.1, 0.2, 0.5\}$. The comparison is shown in Supplementary Fig. S6.

For a single quantum dot, decreasing the timestep systematically reduces the deviation from the reference solution obtained with QmeQ, but also increases runtime. The results for $\delta t = 0.1$ already agree closely with the reference currents while maintaining a reasonable computational cost. Larger timesteps lead to increasing deviations due to integration errors. This correlation was not observed in the four-dot array. In this case, the largest tested timestep, $\delta t = 0.5$, showed closer agreement over part of the parameter range, although all timesteps converge to nearly the same current at $\Gamma = 1$. Overall, the timestep dependence is weaker for four quantum dots than for a single quantum dot. We therefore selected $\delta t = 0.1$ for all production simulations, since it offered the best tradeoff between runtime performance and agreement.

\subsection*{Memory scaling}
The \textit{Lindblad} kernel used in the simulations of QmeQ has a scaling of $64\times\left(2\frac{(2n)!}{(n!)^2} - 2^n\right)$, where $n$ is the number of single-particle states. The TJM MPS memory scaling (to store $N_\text{traj}$ MPS trajectories) is $\mathcal{O}(N_\text{traj}Ld\chi^2_\text{max})$, where $\chi_\text{max}$ is the maximum bond dimension, and $L$ is the system size.

\subsection*{Computational environment}
The simulations were performed on a single high-performance computing node equipped with two AMD EPYC Rome 7H12 processors (128 CPU cores total, 2.6 GHz).

%% file: supplementary_information.tex
\newpage
\setcounter{equation}{0}
	\setcounter{section}{0}
	\setcounter{figure}{0}
	\setcounter{table}{0}
	\setcounter{page}{1}
	\makeatletter
	\renewcommand{\theequation}{S\arabic{equation}}
	\renewcommand{\thefigure}{S\arabic{figure}}
	\renewcommand{\thesection}{S\arabic{section}}

\section*{Supplementary Information}

\subsection*{Matrix-product states}
Matrix-product states (MPS) provide an efficient representation of many-body wavefunctions with limited entanglement. In this work, the system wavefunction is represented as an MPS and the Hamiltonian as a matrix-product operator (MPO). The TDVP algorithm is then used to evolve the MPS in time within a variational manifold of fixed bond dimension.

\subsection*{Time--Dependent Variational Principle (TDVP)}
For the unitary part of the evolution, described by the operator $\mathcal{U}[\delta t]$, the tensor jump method evolves the tensor network (MPS) using the time-dependent variational principle (TDVP) for matrix product states. TDVP approximates the time-dependent Schr\"odinger equation by projecting the evolution vector $-i H |\Psi\rangle$ onto the tangent space of the MPS manifold at $|\Psi\rangle$ before carrying out the time evolution~\cite{paeckel_time-evolution_2019,haegeman_unifying_2016,haegeman_time-dependent_2011}.

More specifically, a dynamic TDVP scheme is employed that combines one-site TDVP (1TDVP) and two-site TDVP (2TDVP). In 1TDVP, the tangent space projector acts on a single MPS tensor $M_l$ at a time. The bond dimensions are kept fixed, and the method yields a strictly unitary evolution within the MPS manifold. The update of the site tensor $M_l$ and the bond matrix $C_l$ is
given by
\begin{equation}
    M_l(t+\delta t) = e^{-i H^{\mathrm{eff}}_l \delta t}\, M_l(t),
    \label{eq:1tdvp_site}
\end{equation}
\begin{equation}
    C_l(t+\delta t) = e^{+i \tilde{H}^{\mathrm{eff}}_l \delta t}\, C_l(t),
    \label{eq:1tdvp_bond}
\end{equation}
where $H^{\mathrm{eff}}_l$ denotes the effective local Hamiltonian obtained from the TDVP projection.

In the two-site TDVP formulation (2TDVP) we instead evolve a pair of neighboring tensors. This requires contracting the site and bond tensors on sites $l$ and $l+1$ into a two-site tensor $N_{l,l+1}$, applying the local time evolution operator, and then splitting the updated tensor again via a single singular-value decomposition (SVD). Schematically,
\begin{equation}
    N'_{l,l+1} = e^{-i H^{\mathrm{eff}}_{l,l+1} \delta t}\, N_{l,l+1} \xrightarrow{\mathrm{SVD}} M'_l\, C'_l\, M'_{l+1},
\end{equation}
where singular values smaller than a truncation threshold $\varepsilon_\mathrm{SVD}$ are discarded. In this way the bond dimension on link $l$ can grow adaptively in order to control the truncation error.

The dynamic TDVP strategy starts from 2TDVP and allows the bond dimensions to increase up to a prescribed maximum $\chi_\mathrm{max}$. Once all bonds reach $\chi_\mathrm{max}$, we switch to 1TDVP and keep the bond dimensions fixed for the remainder of the simulation, which limits the computational cost while still capturing the initial growth of entanglement.

\subsection*{Numerical Master Equation Solver}\label{subsec:det_method}

The dynamics of an open quantum system governed by a Lindblad master equation are generated by the Liouvillian (or Lindbladian) superoperator $\mathcal{L}$ acting on the density matrix:
\begin{equation}
\frac{\partial}{\partial t}\rho(t) = \mathcal{L}[\rho(t)].
\end{equation}
For a time-independent $\mathcal{L}$, the formal solution is
\begin{equation}
    \rho(t) = e^{\mathcal{L} t}\rho(0).
\end{equation}

\noindent In the simplest case, the Liouvillian can be decomposed as
\begin{equation}
\mathcal{L}[\rho] = -\frac{i}{\hbar}[H_S,\rho] + \mathcal{D}[\rho],
\end{equation}
where $H_S$ denotes the system Hamiltonian and $\mathcal{D}$ collects the dissipative contributions. For Markovian open quantum systems, $\mathcal{D}$ is typically written in Gorini--Kossakowski--Sudarshan--Lindblad (GKSL) form (also referred to as the Lindblad master equation)~\cite{campaioli_quantum_2024}:
\begin{equation}
    \begin{split}
    \mathcal{L}[\rho] &= -\frac{i}{\hbar}[H_S, \rho]\\
                      &+ \sum_{m=1}^{k} \gamma_m \left( L_m \rho L_m^\dagger - \frac{1}{2}\{L_m^\dagger L_m, \rho \} \right).
    \end{split}
    \label{eq:lindblad}
\end{equation}
The first term, $-\frac{i}{\hbar}[H_S,\rho]$, describes the unitary component of the dynamics. Throughout this work, we set $\hbar = 1$. The second term captures the non-unitary evolution induced by the environment, where $L_m$ are jump (Lindblad) operators and $\gamma_m$ are the corresponding dissipation rates. The operators $L_m$ need not be Hermitian and encode the different environmental noise channels acting on the system.

For numerical solution, deterministic approaches to the Lindblad master equation~\eqref{eq:lindblad} are commonly formulated in Liouville space, where the density matrix is vectorized and the Liouvillian acts as a linear operator. For a system of $N$ quantum dots with local dimension $d$, the Hilbert-space dimension is $D=d^{N}$, such that the density matrix has $D^2$ elements and the Liouvillian acts on a $D^2$-dimensional operator space. This exponential scaling ($\propto d^{2N}$) limits direct simulation of non-unitary dynamics to small systems~\cite{manzano_short_2020}. Although explicit storage of the Liouvillian would require up to $\mathcal{O}(D^4)$ entries in the dense case, practical implementations typically exploit locality and sparsity and apply $\mathcal{L}$ without constructing it explicitly.

As a reference implementation of this class of solvers, we use QmeQ~\cite{qmeq}, an open-source Python package for transport calculations in quantum-dot systems based on several approximate master-equation approaches, including the Pauli, first-order Redfield, first- and second-order von Neumann, and Lindblad formalisms. Because QmeQ operates in the full many-body Fock space of the central region, its computational cost also scales exponentially with the number of dots, restricting its use to relatively small systems. We therefore benchmark our results against QmeQ in parameter regimes where its underlying approximations are valid.

\subsection*{Quantum Trajectories Method}\label{subsec:qt_method}
The quantum trajectories (Monte Carlo wave-function, MCWF) method is an alternative to the deterministic density-matrix approach, which which represents the open-system dynamics by an ensemble of stochastic pure-state realizations $\{|\psi_i(t)\rangle\}$. Ensemble averages over trajectories reproduce the solution of the Lindblad master equation~\cite{sander2025, molmer_monte_1993, schollwock_density-matrix_2011, campaioli_quantum_2024, benenti_charge_2009, landi_current_2024}.

Given the system Hamiltonian $H_S$ and jump operators $\{L_m\}$ with rates $\{\gamma_m\}$ as in Eq.~\eqref{eq:lindblad}, the no-jump evolution is generated by the non-Hermitian effective Hamiltonian
\begin{equation}
    H_\mathrm{eff} = H_S - \frac{i}{2}\sum_{m=1}^{k}\gamma_m L_m^\dagger L_m .
    \label{eq:heff}
\end{equation}
Between quantum jumps, a trajectory is propagated for a time step $\delta t$ as
\begin{equation}
    |\tilde{\psi}(t+\delta t)\rangle = e^{-i H_\mathrm{eff}\delta t}\,|\psi(t)\rangle ,
\end{equation}
where $|\tilde{\psi}\rangle$ is generally unnormalized. The norm loss determines the total jump probability,
\begin{equation}
    \delta p(t) = 1-\langle \tilde{\psi}(t+\delta t)|\tilde{\psi}(t+\delta t)\rangle
    \simeq \sum_{m=1}^{k}\delta p_m(t),
\end{equation}
with channel-resolved probabilities (for $\delta t$ sufficiently small such that $\delta p\ll 1$)
\begin{equation}
    \delta p_m(t) = \gamma_m \langle \psi(t)|L_m^\dagger L_m|\psi(t)\rangle\,\delta t.
\end{equation}

To sample the stochastic evolution, a random number $\epsilon\in[0,1)$ is chosen. If $\epsilon<\delta p(t)$, a jump occurs; the channel $m$ is selected with probability $\delta p_m(t)/\delta p(t)$ and the state is updated according to
\begin{equation}
    |\psi(t+\delta t)\rangle = \frac{L_m|\psi(t)\rangle}{\sqrt{\langle \psi(t)|L_m^\dagger L_m|\psi(t)\rangle}} .
    \label{eq:jump_update}
\end{equation}
Otherwise no jump occurs and the state is renormalized,
\begin{equation}
    |\psi(t+\delta t)\rangle = \frac{|\tilde{\psi}(t+\delta t)\rangle}{\sqrt{1-\delta p(t)}} .
\end{equation}

Observables are estimated by averaging expectation values over $N_\mathrm{traj}$ independent trajectories,
\begin{equation}
    \langle A\rangle(t) \approx \frac{1}{N_\mathrm{traj}}\sum_{i=1}^{N_\mathrm{traj}}\langle \psi_i(t)|A|\psi_i(t)\rangle ,
    \label{eq:mcwf_expect}
\end{equation}
with statistical uncertainty decreasing as $N_\mathrm{traj}^{-1/2}$.

This kind of approach avoids explicit density-matrix propagation and reduces the memory footprint from $\mathcal{O}(D^2)$ to $\mathcal{O}(D)$ per realization, where $D=d^N$ is the many-body Hilbert-space dimension for $N$ sites of local dimension $d$ (e.g., $D=2^N$ for $N$ qubits). The computational cost per trajectory nevertheless remains exponential in $N$, and overall runtime additionally depends on the number of trajectories required for convergence.

\subsection*{Additional benchmark parameter sweeps}

\begin{figure}[H]
    \centering
    \includegraphics[width=0.75\linewidth]{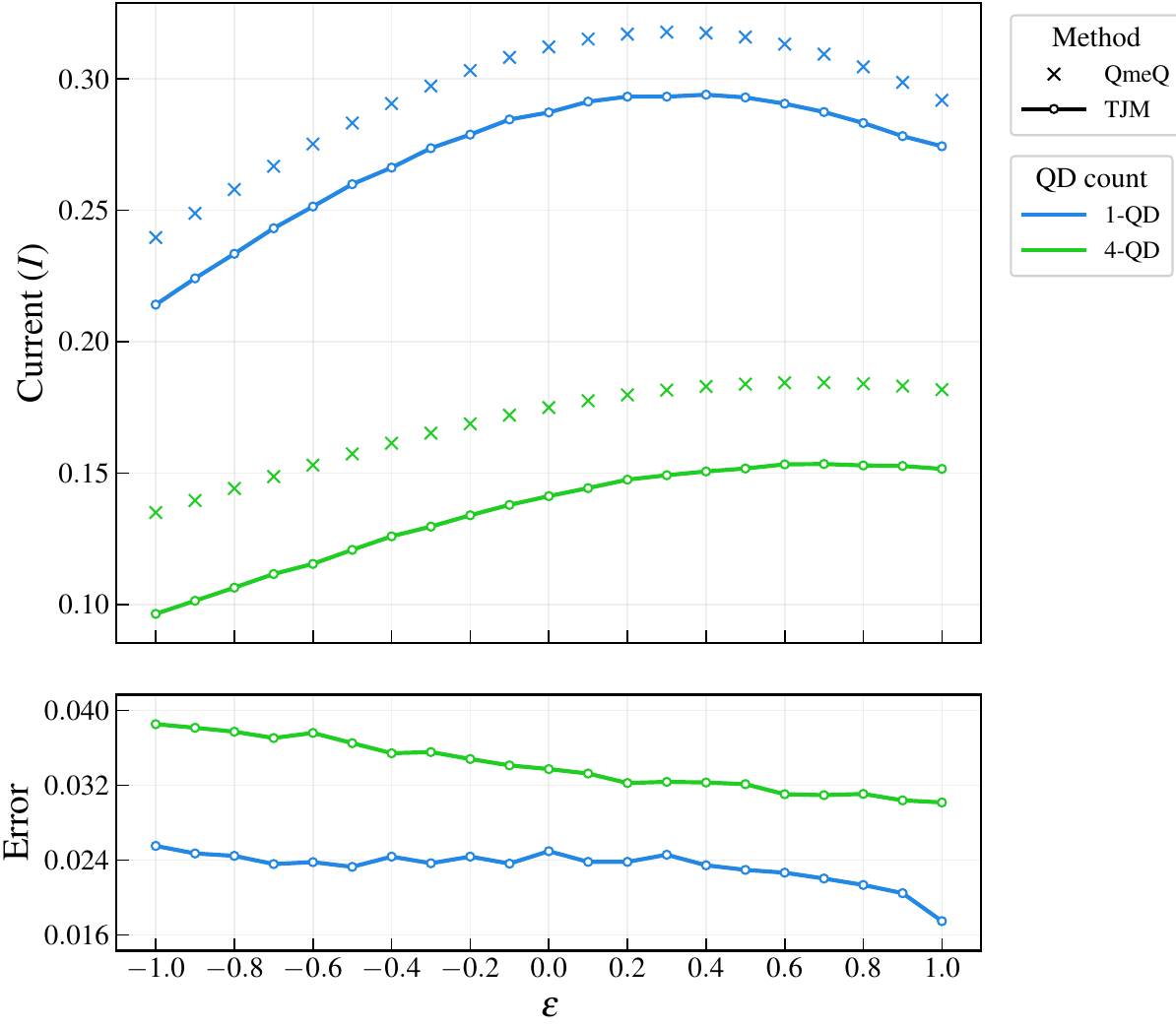}
    \caption{Current comparison between TJM and QmeQ for sweeps in onsite energy.}
\end{figure}

\begin{figure}[H]
    \centering
    \includegraphics[width=0.75\linewidth]{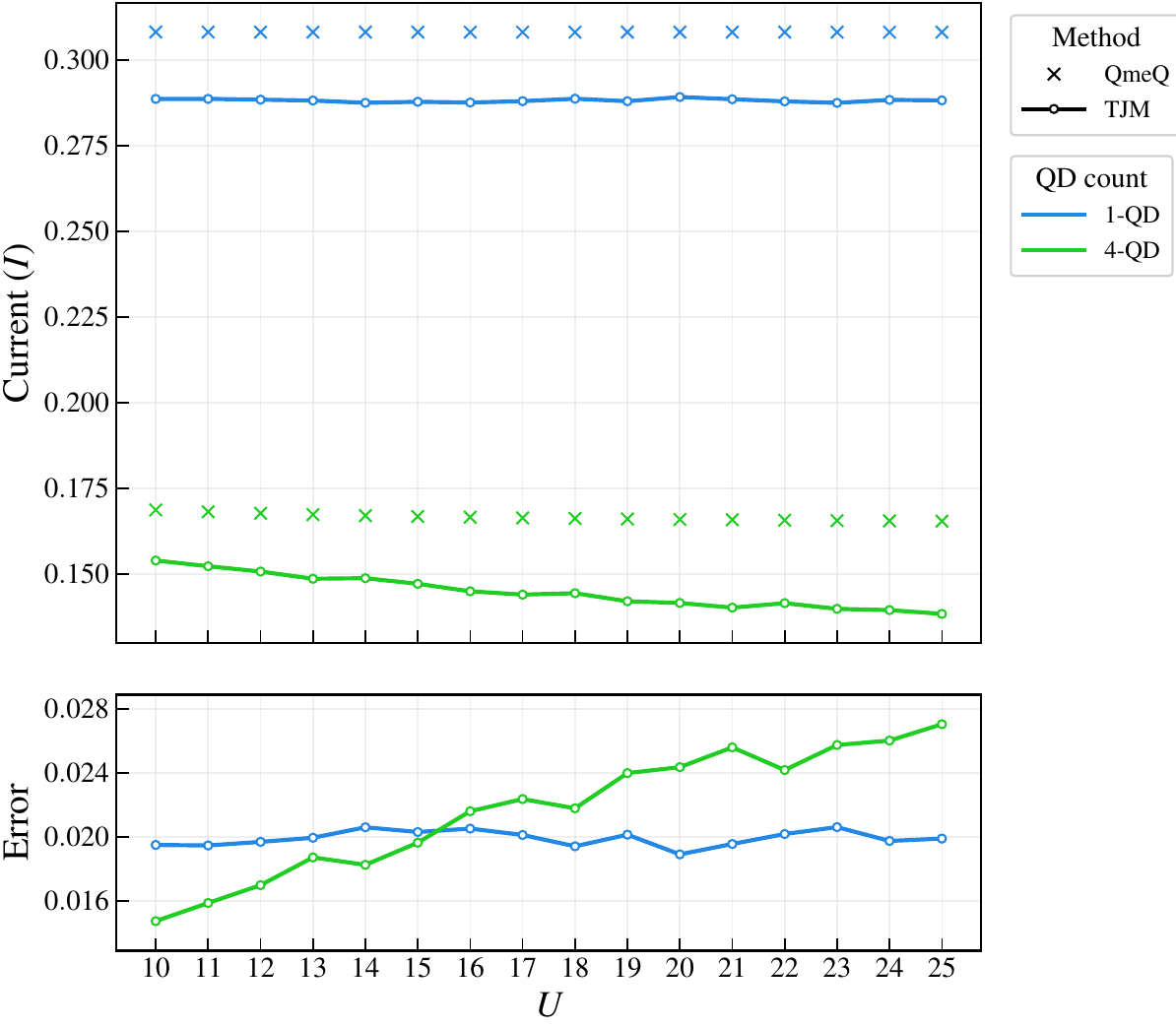}
    \caption{Current comparison between TJM and QmeQ for sweeps in coulomb blockade.}
\end{figure}

\begin{figure}[H]
    \centering
    \includegraphics[width=0.75\linewidth]{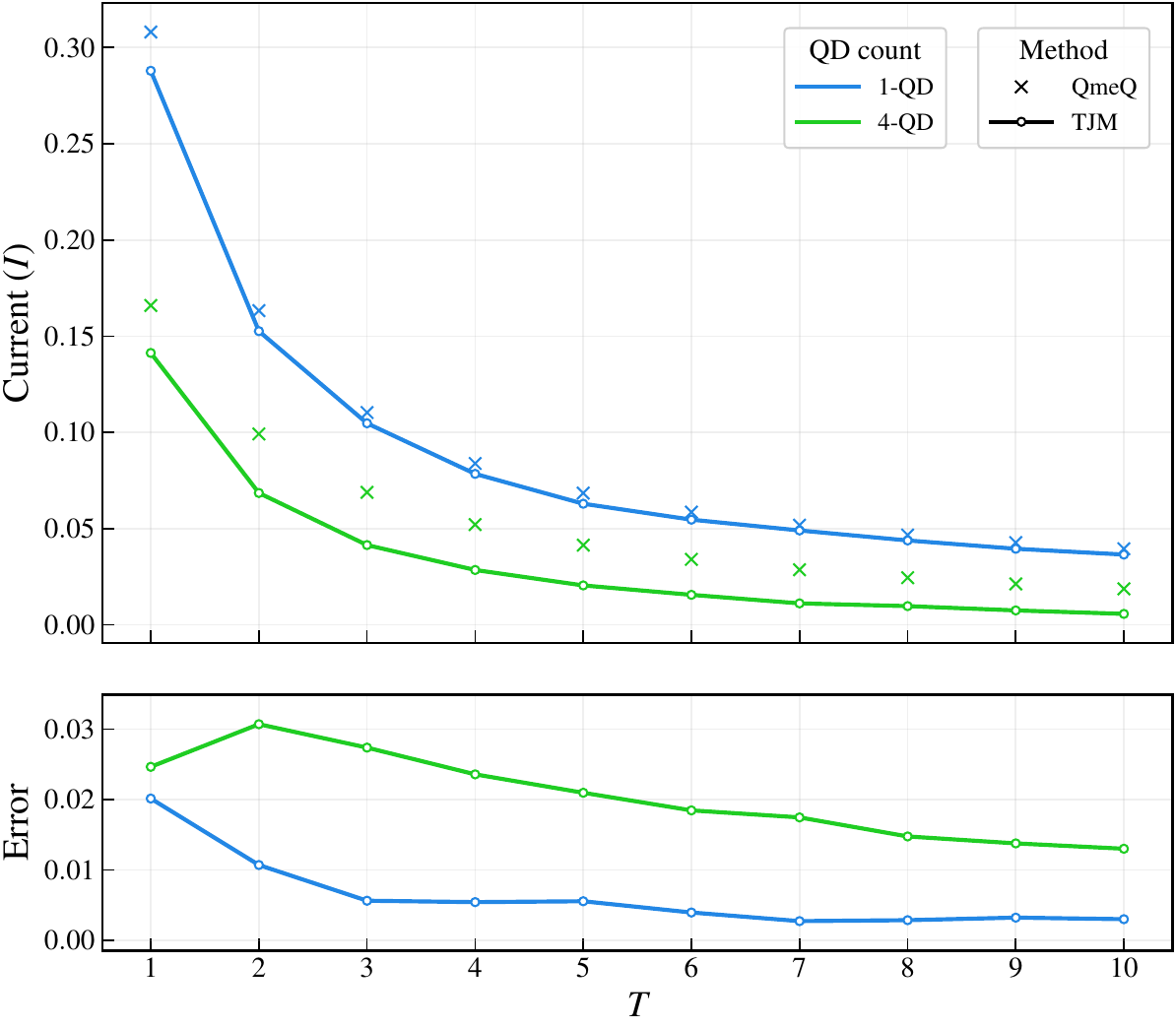}
    \caption{Current comparison between TJM and QmeQ for sweeps in temperature.}
\end{figure}

\subsection*{Trajectory convergence}
\begin{figure}[H]
    \centering
    \includegraphics[width=\textwidth]{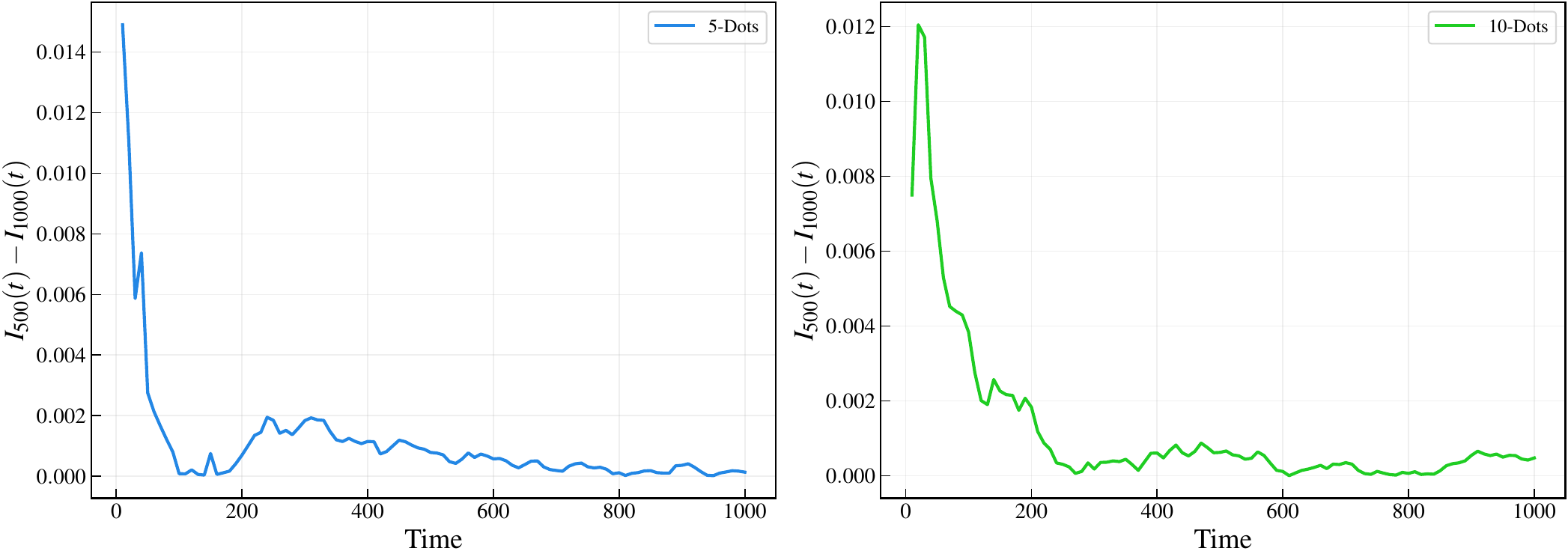}
    \caption{Difference between current traces obtained with 500 and 1000 trajectories for representative 5-dot and 10-dot systems. The late-time deviation remains below $10^{-3}$.}
\end{figure}

\subsection*{Bond-dimension convergence}
\begin{figure}[H]
    \centering
    \includegraphics[width=\linewidth]{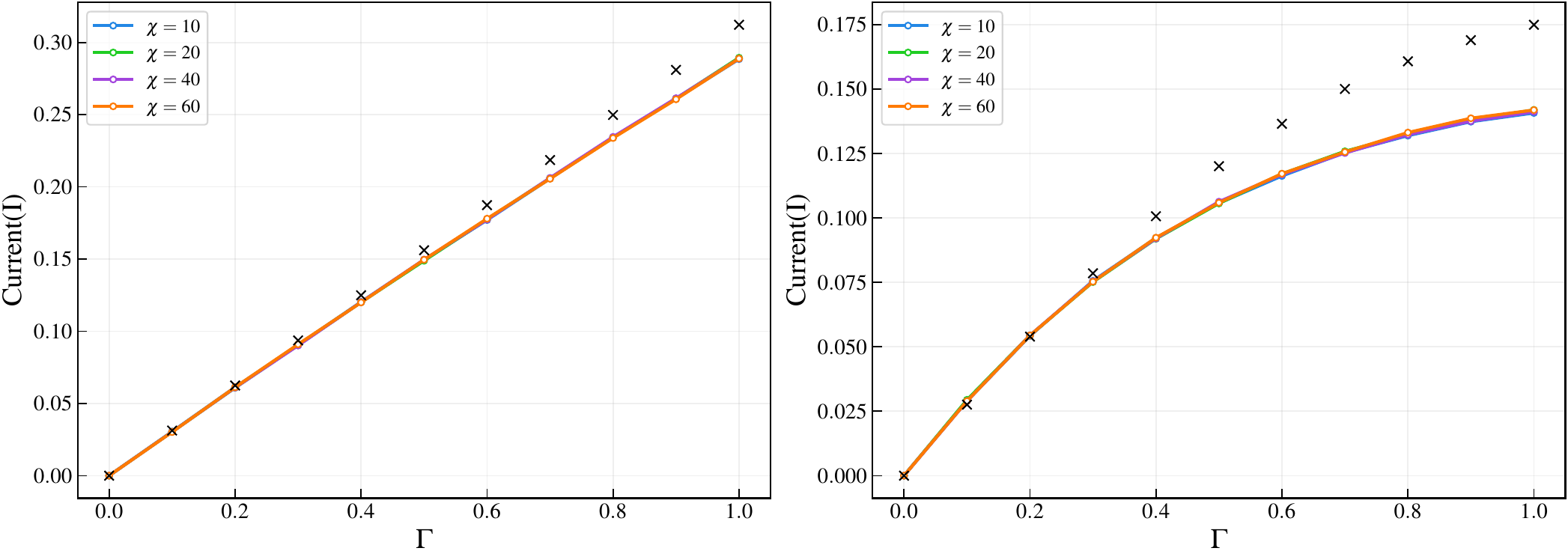}
    \caption{Current traces for different maximum bond dimensions. Current $I$ versus $\Gamma$ computed with the TJM for $\chi_\mathrm{max} \in \{10, 20, 40, 60\}$, for systems containing one quantum dot (left) and four quantum dots (right). Black crosses show the corresponding QmeQ values. Only minor variation is observed across the tested $\chi_\mathrm{max}$ values.}
\end{figure}

\subsection*{Timestep convergence}
\begin{figure}[H]
    \renewcommand\thesubfigure{\textbf{\alph{subfigure}}}
    \captionsetup[subfigure]{justification=raggedright,singlelinecheck=false,labelformat=simple}

    \begin{subfigure}[b]{\linewidth}
        \caption{}
        \centering
        \includegraphics[width=\linewidth]{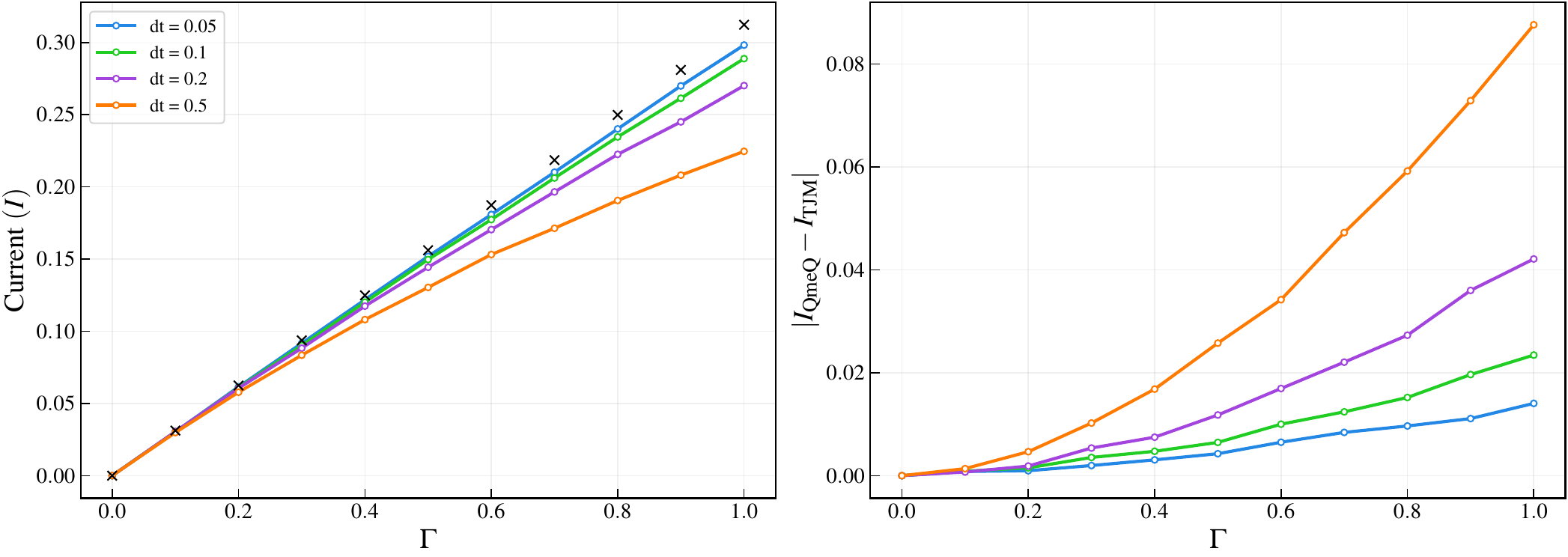}
    \end{subfigure}
    \vfill
    \begin{subfigure}[b]{\linewidth}
        \caption{}
        \centering
        \includegraphics[width=\linewidth]{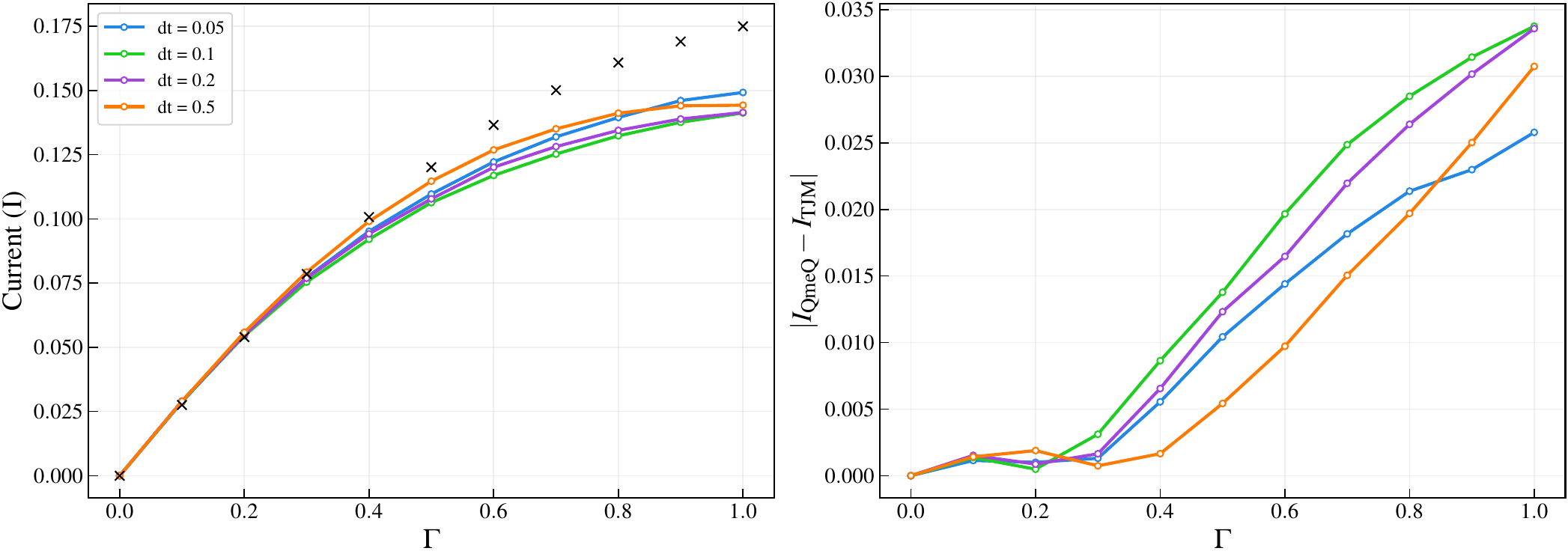}
    \end{subfigure}
    \caption{(a) Current $I$ as a function of $\Gamma$ for a single quantum dot, computed using TDVP timesteps $\delta t = 0.05, 0.1, 0.2, 0.5$, together with the corresponding QmeQ results (black crosses). The right-hand panel shows the absolute deviation $|I_\mathrm{QmeQ}-I_\mathrm{TJM}|$. Smaller timesteps generally yield closer agreement with QmeQ. (b) The same quantities for a four-dot array. In this case, the largest tested timestep, $\delta t = 0.5$, shows closer agreement over part of the parameter range, although all timesteps converge to nearly the same current at $\Gamma = 1$. Overall, the timestep dependence is weaker for four quantum dots than for a single quantum dot.}
\end{figure}